\documentclass[useAMS,usenatbib]{mn2e}

\long\def\symbolfootnote[#1]#2{\begingroup%
\def\thefootnote{\fnsymbol{footnote}}\footnote[#1]{#2}\endgroup} 
\usepackage{graphicx}
\title{Integral field spectroscopy of type-II QSOs at z$=$0.3-0.4}
\author[A. Humphrey et al.]{A. Humphrey$^{1,2}$, M. Villar-Mart{\'{\i}}n$^{3}$, S. F. S\'anchez$^{4,5,6}$, A. Mart{\'{\i}}nez-Sansigre$^{7,8}$,\newauthor
R. Gonz\'alez Delgado$^{3}$, E. P\'erez$^{3}$, C. Tadhunter$^{9}$, M.-A. P\'erez-Torres$^{3}$\\
$^{1}$Korea Astronomy and Space Science Institute, 61-1 Hwaam-dong, Yuseong-gu, Daejeon, 305-348, Republic of Korea\\
$^{2}$Instituto Nacional de Astrof\'{\i}sica, \'Optica y Electr\'onica (INAOE), Aptdo. Postal 51 y 216, 72000 Puebla, Pue., Mexico (ajh@inaoep.mx)\\
$^{3}$Instituto de Astrof\'{\i}sica de Andaluc\'{\i}a (CSIC), Aptdo. 3004, 18080 Granada, Spain\\
$^{4}$Centro Astron\'omico Hispano Alem\'an, Calar Alto, (CSIC-MPG), C/Jes\'us Durb\'an Rem\'on 2-2, E-04004 Almeria, Spain\\
$^{5}$Centro de Estudios de F\'\i sica del Cosmos de Aragon (CEFCA), C/General Pizarro 1, 3º, E-41001 Teruel, Spain\\
$^{6}$Fundaci\'on Agencia Aragonesa para La Investigaci\'on y el Desarrollo (ARAID)\\
$^{7}$Astrophysics, Department of Physics, University of Oxford, Keble Road, Oxford OX1 3RH, UK \\
$^{8}$Institute of Cosmology and Gravitation, University of Portsmouth, Dennis Sciama Building, Burnaby Road, Portsmouth, \\ PO1 3FX, United Kingdom \\
$^{9}$Department of Physics and Astronomy, University of Sheffield, Hounsfield Road, Sheffield S3 7RH, UK}

\date{Accepted 2010 June 16.
      Received 2010 May 19;
      in original form 2009 August 17}

\pagerange{\pageref{firstpage}--\pageref{lastpage}}
\pubyear{2010}

\begin{document}
\topmargin -0.5in
\maketitle

\label{firstpage}

\begin{abstract}
We present and analyse integral-field observations of six type-II QSOs with z=0.3-0.4, selected from the Sloan Digital Sky Survey (SDSS).  Two of our sample are found to be surrounded by a nebula of warm ionized gas, with the largest nebula extending across 8\arcsec (40 kpc).  Some regions of the extended nebulae show kinematics that are consistent with gravitational motion, while other regions show relatively perturbed kinematics: velocity shifts and line widths too large to be readily explained by gravitational motion.  We propose that a $\sim 20$ kpc $\times 20$ kpc outflow is present in one of the galaxies.  Possible mechanisms for triggering the outflow are discussed.  In this object, we also find evidence for ionization both by shocks and the radiation field of the AGN.  
\end{abstract}


\section{Introduction}
Powerful active galactic nuclei present unique opportunities for investigating the formation and evolution of massive galaxies, as well as the interplay between the active galactic nucleus (AGN) and its environment.  Type-II (i.e., obscured) radio-quiet quasars (QSOs hereinafter) are of particular interest for this.  Their high luminosities should afford their detection across large ranges in redshift, while their lack of powerful radio jets should simplify the interpretation of the observable properties of their host galaxies and environments.  Moreover, most of the growth of super massive black holes occurs in obscured quasars (e.g. Martinez-Sansigre et al. 2005).  In addition, unlike their Type-I (unobscured) counterparts, the optical emission from the host galaxy is unlikely to be overwhelmed by the AGN.  

The existence of Type-II QSOs was required by unification models of active galaxies (e.g. Barthel 1989; Antonucci 1993 and references therein) and, although long sought-after, it was only with the Sloan Digital Sky Survery (SDSS) that these objects were detected in significant numbers (Zakamska et al. 2003).  To date, there have been a number of spectroscopic investigations of Type-II quasars (e.g. Zakamska et al. 2003; Villar-Mart{\'{\i}}n et al. 2008), but these have been relatively superficial and based predominatly on data from the SDSS, focussing primarily on classifying the objects, understanding the nature of the powering mechanism and the obscuring structure, and testing unification models.  

While these studies have provided a wealth of important information, little work has been carried out to characterize the gaseous properties of Type-II quasars.  In this letter we present integral field spectroscopic observations for 6 moderate redshift type II QSOs, 5 of which are radio quiet ($L_{1.4GHz}<10^{31} erg s^{-1} Hz^{-1} Sr^{-1}$).  Our main goals are to investigate the possible existence of extended ionized nebulae around the galaxies, and characterize for the first time their kinematic, morphological and excitation properties.  Throughout this letter, we assume a flat universe with $H_{0}$=71 km $s^{-1}$ Mpc$^{-1}$, $\Omega_{\Lambda}$=0.73 and $\Omega_{m}$=0.27.

\section{Observations and Reduction}
The sample we use in this investigation consists of one radio-loud and five radio-quiet type-II quasars (QSOs) at z$\sim$0.3-0.4.  This redshift range was used to enable us to observe from [NeV]$\lambda$3426 to [OIII] $\lambda$5007, and also to minimise any redshift evolution effects.  All of our sources were selected from the sample of type-II QSOs originally identified by Zakamska et al. (2003) in the SDSS.  The objects were selected to have relatively high [OIII] $\lambda$5007 emission line fluxes, i.e, $\ge 1 \times 10^{-15}$ erg s$^{-1}$ cm$^{-2}$, in order to maximise the probability of detecting extended nebulae.  We also required that source coordinates permit observations from Calar Alto. 

Our observations were taken on 2008 January 8 and 10, and on 2008 July 1 and 2, at the 3.5m telescope of the Calar Alto observatory, Spain, using the Potsdam Multi-Aperture Spectrograph (PMAS:
Roth et al. 2005) in its lens-array mode.  The atmospheric conditions were
variable.  During the January observing run the humidity was high (i.e., non-photometric).  
In the July run the conditions were
photometric. The seeing was relatively stable, ranging between 1.2$\arcsec$ and
1.6$\arcsec$ (or $\sim$6-8 kpc).  The PMAS lens-array comprises a bundle of 256 fibres, coupled to a
square array of 256 lenses, and resulted in contiguous sampling of the sky.  The optics were adjusted to give a spatial scale of 
1$\arcsec$/spaxel (lens) and a field of view of 16$\arcsec$$\times$16$\arcsec$. 
Cross-talk between adjacent fibres is estimated in less than a 3\% when using a pure aperture extraction (see
below).  We used the V300 grating, covering a spectral range of $\sim$4250-7650\AA, wherein
a number of interesting emission lines lie.  The full-width at half maximum 
(FWHM) of the instrumental profile, measured from the strong sky line [OI] $\lambda$5577\AA, was 7.4\AA\ $\pm$0.5\AA.  Our observations are summarized in Table 1.  

\begin{table*}
\caption{Summary of the observations.  Column 3 lists the radio type: RQ=radio quiet ($L_{1.4GHz}<10^{31} erg s^{-1} Hz^{-1} Sr^{-1}$); RL=radio loud ($L_{1.4GHz}\ge 10^{31} erg s^{-1} Hz^{-1} Sr^{-1}$).  Column 7 gives the 1$\sigma$ sensitivity of the observations relative to that of SDSS J084041.08+383819.8.  Column 8 gives the mean surface brightness of the [OIII] $\lambda$5007 line, in units of $10^{-16}$ erg s$^{-1}$ cm$^{-2}$ arcsec$^{-2}$, at a radius of 3\arcsec from the peak of the optical continuum.  Columns 9 lists the FWHM of[OIII] $\lambda$5007 after deconvolution from the instrumental profile (units of km s$^{-1}$), in the spaxel corresponding to the continuum flux maximum.}
\begin{tabular}{lllllllll}        
\hline                 
Target & z & Type & Date & Exp. (s) & Notes &  $\sigma$/$\sigma_0$ & SB$_{[OIII]}$ & FWHM \\
\hline
SDSS J084041.08+383819.8 & 0.313 & RQ & 10/01/08 & 8$\times$1800 & high humidity & 1.0 & 1.6$\pm$0.1 & 1330$\pm$130 \\
SDSS J092014.11+453157.3 & 0.402 & RQ & 08/01/08 & 2$\times$1800 & cloudy        & 2.0 & $\le$0.6    & 700$\pm$100 \\
                    &       &    & 10/01/08 & 2$\times$1800 & cloudy             & &             &              \\
SDSS J155059.30+395029.5  & 0.347 & RQ & 01/07/08 & 6$\times$1200 & photometric  & 1.8 & 4.8$\pm$0.2 & 920$\pm$90 \\
SDSS J162545.11+310632.2 & 0.379 & RL & 02/07/08 & 6$\times$1200 & clear         & 3.6 & $\le$0.9    & 540$\pm$40 \\
SDSS J172603.07+602115.6 & 0.333 & RQ & 01/07/08 & 6$\times$1200 & photometric   & 2.2 & $\le$0.3    & 900$\pm$100 \\
SDSS J173938.64+544208.6 & 0.384 & RQ & 02/07/08 & 7$\times$1200 & clear         & 2.8 & $\le$0.2    & 660$\pm$70 \\
\hline                                   
\end{tabular}
\end{table*}

Data reduction was performed using R3D (S\'anchez 2006), in combination with
IRAF packages and E3D (S\'anchez 2004). The reduction
consists of the standard steps for fibre-based integral-field spectroscopy (S\'anchez 2006), 
including bias and flat-field corrections, tracing and extraction of spectra, 
geometrical distortion correction, and calibration of the wavelength and flux scales.  
Finally, the spectra were assembled into a data-cube.  We made no correction
for galactic extinction, because it is low for all sources; uncertainties introduced 
by not applying the correction are expected to be substantially smaller than those introduced by 
the flux calibration.

\section{Analyses}

\subsection{Emission line images}
In order to examine the spatial distribution of the line emitting gas associated with the galaxies, we have reconstructed images of the [OIII] emission doublet and, where the signal-to-noise ratio was sufficient, [OII].  First, we summed the line flux along the dispersion axis of the data cube, to produce line plus continuum images.  For each combination of object and emission line, we used a fixed wavelength range selected to include the wings of the line, and also to allow for any velocity shifts between lenses, corresponding to an image bandwidth of between 110\AA~ and 220\AA.  In the case of the [OIII] doublet, we included both the 4959\AA~ line and the 5007\AA~ line.  Next, we reconstructed images for the continuum emission from a spectral region nearby to the emission line, and with the same wavelength bandwidth as the relevant line plus continuum image.  These were subtracted from the line plus continuum images, to produce pure line images.  The reconstructed emission line images are shown in Fig. 1.  In each image, the lowest contour level represents the 3$\sigma$ detection limit.  We do not detect spatially extended line emission from SDSS J092014.11+453157.3, SDSS J162545.11+310632.2, SDSS J172603.07+602115.6 or SDSS J173938.64+544208.6 (see column 8 of Table 1).  This is predominantly due to the faintness (or lack of) extended line emission from these galaxies.  We find no clear trend between detection/non-detection of extended nebulae and the properties of the emission form the central spaxel or the radio properties (Table 1; c.f. Christensen et al. 2006; Husemann et al. 2008).  The four galaxies without detections of extended nebulae will not be considered further in this letter.  For SDSS J155059.30+395029.5 and SDSS J084041.08+383819, we also show [OII]/[OIII] maps in Figs. 2 and 5, respectively.

\begin{figure}
\includegraphics{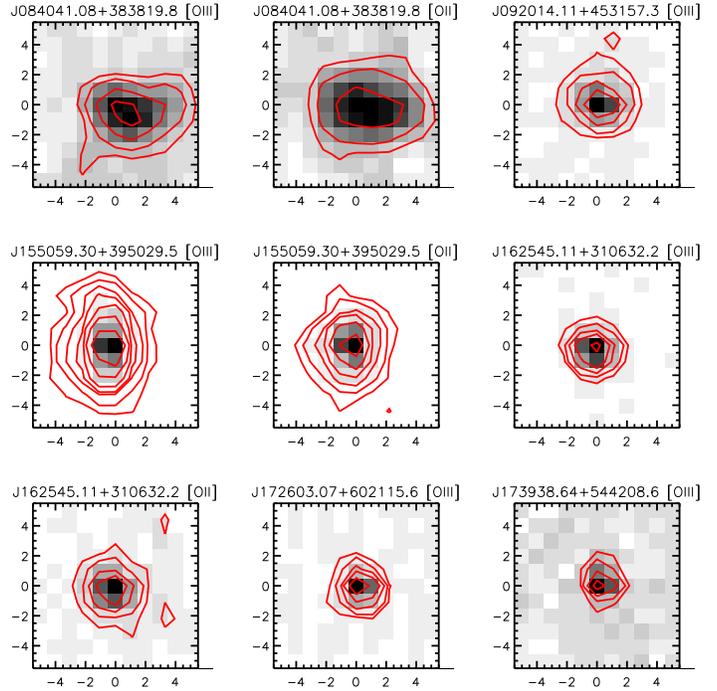}

\vspace{3.585in}\caption{Reconstructed emission line images for our sample.  North is at the top and East is to the left.  The axes give the spatial offset, in arcsec, from the continuum peak.  The contours are plotted for 1, 2, 4, 6.7, 13, 25 ,50 $\times$ 3$\sigma$ when possible.  The line emission from SDSS J084041.08+383819 and SDSS J155059.30+395029.5 is spatially extended on scales of several tens of kpc; their significant spatial variations in kinematic properties and in line ratios confirms their spatial extensions (Figs. 2 and 5).  In the rest of the sample, we find no evidence spatially extended line emission at the resolution of our observations ($\sim$6-8 kpc).}
\end{figure}

\subsection{Velocity and FWHM maps}
Maps of the FWHM and line of sight velocity were produced for [OIII] $\lambda$5007, typically the brightest emission line in these spectra.  We fitted the spectral region around [OIII] $\lambda$5007 with a single Gaussian velocity profile atop a linear continuum, making use of MPFIT and other routines from the Markwardt IDL library.  We resticted ourselves to fit only those lenses where the [OIII] $\lambda$5007 velocity profile had a peak signal-to-noise ratio of $\ge$7.  Although the kinematics are complex in some lenses, we have nevertheless fitted {\it single} Gaussians because our goal is obtain a general characterisation of the kinematic properties.  FWHM values have been corrected for instrumental broadening.  FWHM and $\Delta$v have typical 1$\sigma$ uncertainties of $\sim$100 km s$^{-1}$.  

\section{Results and discussion}

\subsection{SDSS J155059.30+395029.5}
Although this source meets our criteria for being radio-quiet, it does emit a significant radio flux density.  The flux density measurement from the FIRST survey (Becker, White \& Helfand 1995), which has a spatial resolution of 5\arcsec (24 kpc), is 10.6$\pm$0.5 mJy.  From the NVSS survey (Condon et al. 1998), with a resolution of 45\arcsec (220 kpc), the flux density is measured to be 13.1$\pm$0.6 mJy.  The fact that the NVSS measurement is significantly higher suggests that there is emission on large spatial scales, and this is almost certainly emission from radio jets.  

This object is associated with a spatially extended emission line nebula (Fig 1), with an integrated [OIII] $\lambda$5007 luminosity of 1.5$\pm$0.2$\times 10^{43}$ erg s$^{-1}$.  The [OIII] $\lambda\lambda$4959,5007 and [OII] $\lambda$3727 doublets extend across 8\arcsec (40 kpc) and 6\arcsec (30 kpc), respectively.  H$\beta$ and [NeV] $\lambda$3426 are also spatially extended.  The nebula is elongated approximately along the North-South axis, and shows an excess of line emission on East side of the emission peak relative to the West.  

The kinematic properties vary significantly across the nebula: 130$\le$FWHM$\le$1070 km s$^{-1}$ and -20$\le$$\Delta$v$\le$310 km s$^{-1}$ (Fig. 2).  The outer regions of the nebula have narrower emission lines, i.e.,$\le$420 km s$^{-1}$, consistent with gravitational motion, and the blue wing is weak or undetected therefrom.  We also note that regions with lower FWHM tend to have higher line of sight velocities (Fig. 4).  

The highest FWHM are measured in the spaxel at the spatial zero, and in several spaxels to the immediate North and East thereof (up to 1070 km s$^{-1}$).  In these spaxels, a substantial blue wing is also apparent in the velocity profiles (Fig. 4), with a total detected extent of 4\arcsec$\times$4\arcsec (20 kpc $\times$ 20 kpc).  Since these kinematic properties are not centred on the flux peak of the line and continuum, they are unlikely to be the result of contamination by the central regions of the AGN.  We define a 'detection' of the blue wing as follows.  We perform two separate fits to the line profile: one using a single Gaussian and another using two Gaussians.  We formally 'detect' the blue wing when both of the following conditions are met: (1) the two-Gaussian fit shows an excess of flux on the blue side of the line profile, and (2) the reduced $\chi^2$ is $\ge$16 higher for the two-Gaussian fit than it is for the single Gaussian fit.  

The presence of these highly perturbed kinematics suggests the existence of an outflow (e.g. Jarvis et al. 2003).  In the absence of a detection of photospheric absorption lines, we will assume that the quiescent gas is at the systemic velocity, and that the blue wing is due to an outflow.  Next, we attempt to quantify the observed properties of this outflow.  After summing the spaxels from which we detected the blue wing, we fitted the line profile of [OIII] $\lambda$5007 using two Gaussians (Fig. 4).  We use [OIII] $\lambda$5007 because it is the highest signal to noise line in this spectrum.  From the resulting fit, we take the relatively broader, blueshifted component to represent the outflowing gas: it has a (convolved) FWHM=1007 km s$^{-1}$, is blueshifted from our fiducial velocity zero by $v_{o}$=285 km s$^{-1}$, and has a [OIII] $\lambda$5007 flux of 1.7 $\times$10$^{-14}$ erg s$^{-1}$ cm$^{-2}$.  Scaling this to H$\beta$ using the [OIII] $\lambda$5007 / H$\beta$ ratio measured in the same aperture, we obtain a H$\beta$ flux of 2.7 $\times$10$^{-15}$ erg s$^{-1}$ cm$^{-2}$, which corresponds to L$_{H\beta}$=1.1 $\times$10$^{42}$ erg s$^{-1}$.  While this parameterisation is likely to be a gross oversimplification of reality, it is a necessary one given our limited information about the true geometry and dynamics of the outflow.  Assuming spherical outflow, the gas outflow rate is then

\begin{equation}
\dot{M} = \frac{3 v_{o} L_{H\beta} m_p}{r n h \nu _{H\beta} \alpha_{H\beta}^{eff} }
\end{equation}

where $m_p$ is the proton mass, $r$ is the radius of the outflow, $n$ is the density of the outflowing gas, $h\nu _{H\beta}$ is the energy of an H$\beta$ photon and $\alpha_{H\beta}^{eff}$ is the effective H$\beta$ recombination coefficient (e.g. Dopita \& Sutherland 2003).  For the outflow radius, we adopt the maximum projected spatial extent of the blue wing from the optical nucleus ($\ge$2\arcsec or $\ge$3.0 $\times$10$^{22}$ cm).  In addition, we assume n$\le$1000 cm$^{-3}$, based on the strong detection of the forbidden [OII] doublet; the [SII] $\lambda\lambda$6716,6731 doublet profile also suggests densities lower than 1000 cm$^{-3}$.  Hence, we obtain a very approximate outflow rate of $\ga 2 M_{\odot}$yr$^{-1}$.  

A galactic scale ouflow, as observed in relatively nearby galaxies, is one possible physical explanation.  Such outflows can have spatial scales as large as $\sim$10 kpc, outflow rates of $\sim$1-100 $M_{\odot}$yr$^{-1}$, and are powered by active star formation with rates of $\ga$10 $M_{\odot}$yr$^{-1}$ (e.g. Heckman, Armus \& Miley 1990; Lehnert \& Heckman 1996; Heckman et al. 2000).  If the entire FIRST flux density were due to star-formation, it would imply a star formation rate in excess of 1000 $M_{\odot}$yr$^{-1}$, which would be more than sufficient to power the outflow from SDSS J155059.30+395029.5.  

In addition, we note that kinematically perturbed emission lines have been detected from the extended ionized gas of radio-loud quasars and of powerful radio galaxies, and are generally attributed to the radio jet activity (e.g. Clark et al. 1998).  If the radio emission from SDSS J155059.30+395029.5 is dominated by emission from the AGN, rather than powered by star formation, could such a radio source power the outflow in SDSS J155059.30+395029.5?  Assuming a radio spectral index of 1, and using the NVSS radio flux density, we calculate a radio luminosity of $\sim$4$\times 10^{31}$ erg s$^{-1} $Hz$^{-1}$ Sr$^{-1}$.  Then, using the relation $L_{kin} = 10^{28\pm2}L_{radio}^{0.35\pm0.05}$ from Birzan et al. (2004), we estimate the kinetic jet power to be between 2$\times 10^{37}$ and 1$\times 10^{45}$ erg s$^{-1}$.  For an outflow velocity $v_{o}=285$ km s$^{-1}$, a gas density of 1 cm$^{-3}$ and injection time of 10 Myr, the required energy injection rate is $\sim 10^{43}$ erg s$^{-1}$ (see e.g. equation 2 in Nesvadba et al. 2006), in agreement with the higher end of the range in possible kinetic jet power: thus, the radio source does appear able to power the ionized gas outflow.  

A further possibility is that the outflow is powered by the radiation field of the active nucleus: an outlfow is predicted for quasars accreting near to, or above, the Eddington limit (e.g. King \& Pounds 2003).  In principle, outflows of this kind can escape from the central region of the AGN (King \& Pounds 2003; Pounds et al. 2003a,b) and their spatial extent can be of the order of tens of kpc (Silk \& Rees 1998; King 2003).  Assuming the relationship between the bolometric and [OIII] $\lambda$5007 luminosities $L_{bol} \sim 1000 \times L_{[OIII]}$ (e.g. Zakamska et al. 2003; Richards et al. 2006), the observed [OIII] luminosity of $L_{[OIII]}=$1.5$\pm$0.2$\times 10^{43}$ erg s$^{-1}$ implies $L_{bol} \sim 1.5 \times 10^{46}$ erg s$^{-1}$, which in turn implies a black hole mass of 1.2 $\times 10^{8} M_{\odot}$.  Assuming a typical radiative efficiency of 5 per cent (Martinez-Sansigre \& Taylor 2009), this implies an accretion rate of $\sim5 M_{\odot}$yr$^{-1}$.  King \& Pounds (2003) argue that the outflow rate will be roughly the same as the accretion rate, in which case the outflow rate would be in agreement with the rate of $\ga 2 M_{\odot}$yr$^{-1}$ we estimated for SDSS J155059.30+395029.5.  We note that this kind of outflow might provide a plausible explanation for the perturbed kinematics in some of the extended nebulae associated with radio galaxies and quasars, which heretofore have been assumed to be caused by their powerful radio jets.  

The [OII]/[OIII] ratio varies substantially across the nebula: 0.3$\le$[OII]/[OIII]$\le$1.7.  At x=1\arcsec N (5 kpc N), y=2\arcsec W (10 kpc W), this ratio reaches its maximum, suggesting a relatively low state of ionization.  In addition, the [OIII]/H$\beta$ ratio of 2.1$\pm$0.8 is also consistent with low a ionization state at this position (Fig 3).  The detection of spatially extended [NeV] $\lambda$3426 suggests that the hard radiation field of the active nucleus contributes significantly to the ionization of the nebula.  Interestingly, [NeIII] $\lambda$3869 / [OIII] $\lambda$5007$=$1.0$\pm$0.3, which is unusually high and is consistent with ionization by shocks, but is ten times higher than would be expected for photoionized gas (e.g. Dopita \& Sutherland 1996).  We suggest that shock-ionization is a consequence of the spatially-extended outflow in this source.  

Moreover, in the spaxels where the lines are broadest, we find that the FWHM of H$\beta$ is systematically higher than those of [OIII] $\lambda\lambda$4959,5007 by between 100$\pm$40 and 220$\pm$40 km s$^{-1}$, suggesting that the blue wing / outflow has a lower ionization state than the kinematically quiescent gas.  This is also consistent with the effects of shocks (Clark et al. 1998).  

\subsection{SDSS J084041.08+383819.8}
We measure spatial extents of 5\arcsec (23 kpc) and 6\arcsec (27 kpc) for the [OIII] $\lambda\lambda$4959,5007 and [OII] $\lambda$3727 emission lines, respectively.  The total luminosity of the [OIII] $\lambda$5007 line is 9.8$\pm$1.5$\times 10^{41}$ erg s$^{-1}$.  While the [OII] flux peaks at the position of the optical continuum peak, the [OIII] emission flux reaches its maximum 1\arcsec (5 kpc) W,1\arcsec (5 kpc) S of this position.  The spatial distribution of the line emission is elongated towards the West.  

Fig. 5 shows the FWHM, line of sight velocity, and [OII] $\lambda$3727 / [OIII] $\lambda\lambda$4959,5007 ratio maps.  FWHM ranges between 580 and 1150 km s$^{-1}$, with the highest values being measured at, or adjacent to, the spatial zero.  A prominent red wing is visible in the [OIII] velocity profile in the 3\arcsec$\times$3\arcsec (14 kpc $\times$ 14 kpc) centred on the spatial zero; it is not clear whether these kinematics originate in the extended gas, or are instead due to contamination by emission from the central regions of the AGN.  Line of sight velocity varies between -80 and 230 km s$^{-1}$.  The relatively broad emission lines at some spatial positions reveal the presence of highly perturbed gas in the nuclear region, or close to it.  The [OII] $\lambda$3727 / [OIII] $\lambda\lambda$4959,5007 ratio varies from 0.5-1.1 across the nebula, with lower values closer to the spatial peak of the [OIII] emission.  Overall, the line ratios are consistent with ionization by shocks or the radiation field of an AGN.  

\begin{figure*}
\includegraphics{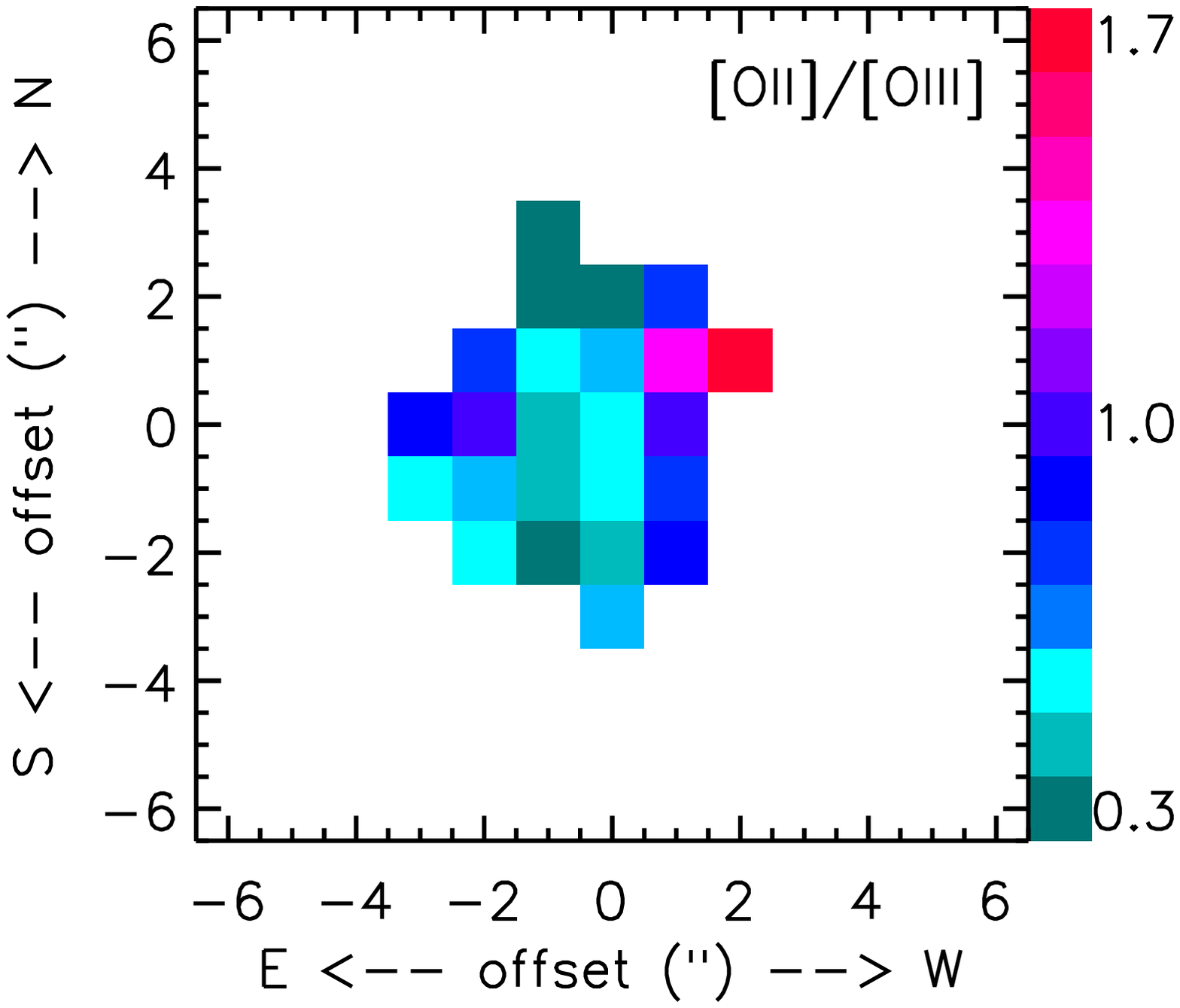}
\includegraphics{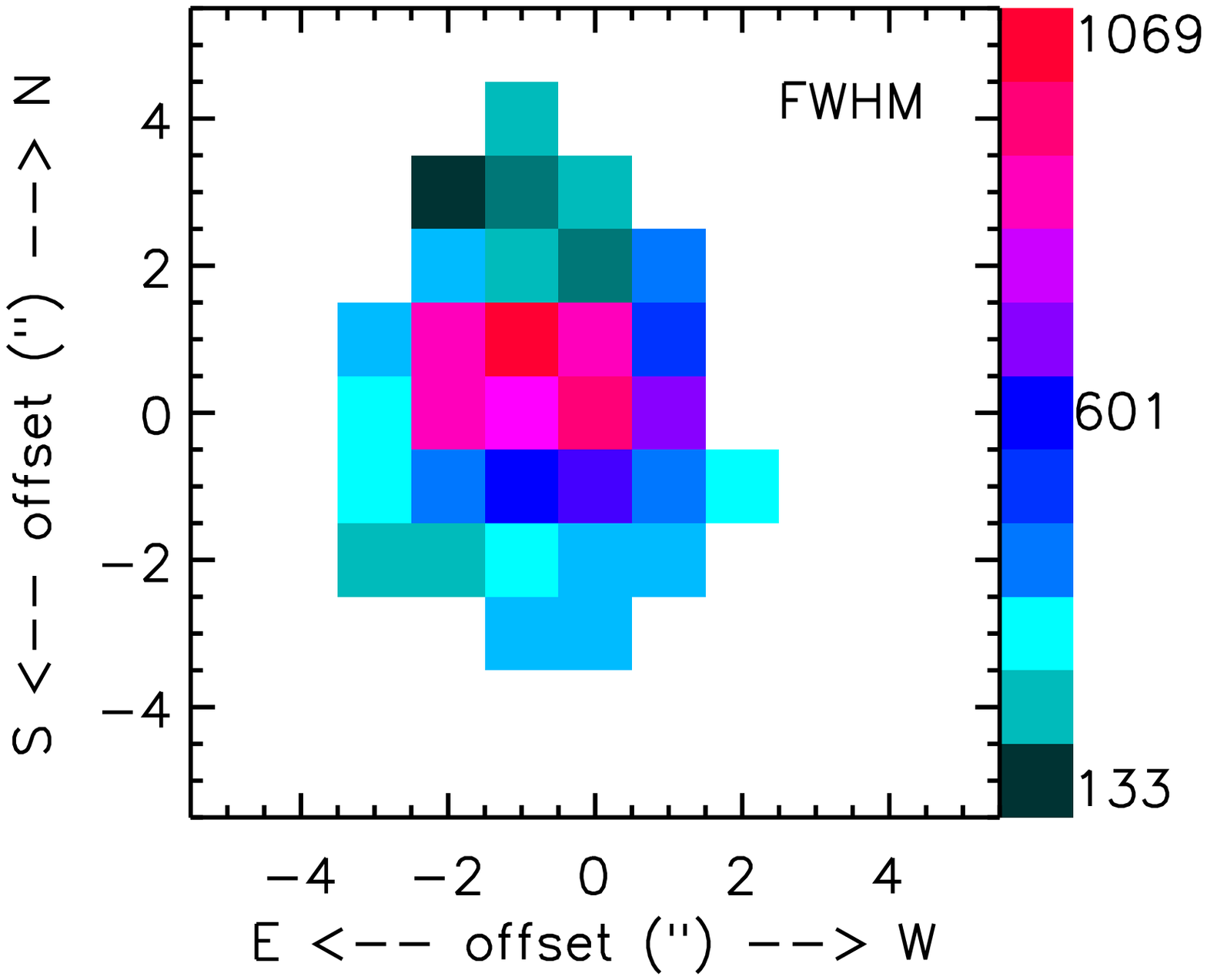}
\includegraphics{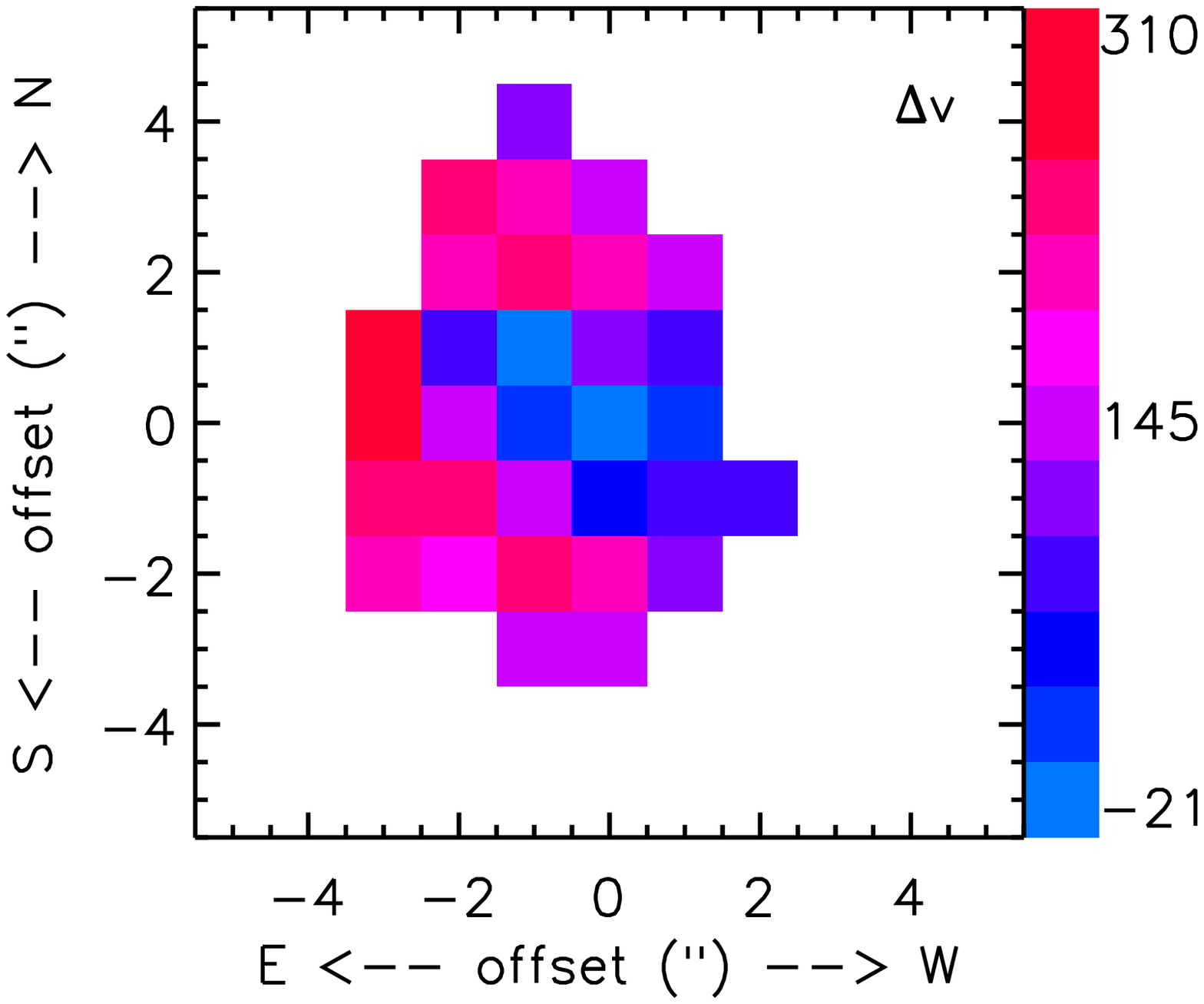}
\vspace{2.1in}\caption{Results from the datacube of SDSS J155059.30+395029.5.  From left to right: the [OII] $\lambda$3727 / [OIII] $\lambda\lambda$4959,5007 map; the [OIII] $\lambda$5007 FWHM map (km s$^{-1}$); the map of the velocity shift of [OIII] $\lambda$5007 (km s$^{-1}$), where we adopt the velocity at the spatial zero as our fiducial velocity zero.  The range in [OII] $\lambda$3727 / [OIII] $\lambda\lambda$4959,5007,FWHM and velocity shift is typical of the spatially extended ionized gas associated with powerful active galaxies (see text).  The total intensity map is similar to that shown in Fig. 2.}
\end{figure*}

\begin{figure}
\includegraphics[scale=0.66,clip=true,trim=1.5cm 14.5cm 0cm 0.8cm]{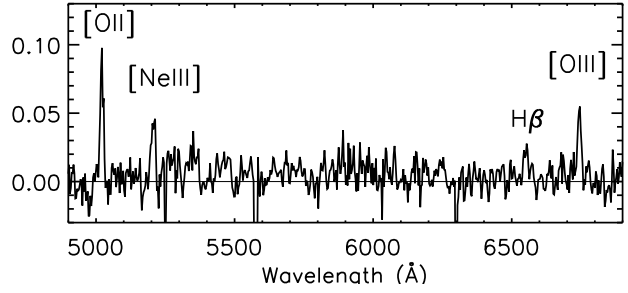}
\caption{Spectrum from the lens at x=1$\arcsec$ (5 kpc) N, y=2$\arcsec$ (10 kpc) W in SDSS J155059.30+395029.5.  Notice the relative weakness of [OIII] $\lambda$5007 compared to [OII], [NeIII] and H$\beta$; [OIII]$\lambda$5007/H$\beta$=2.1$\pm$0.8 and [NeIII]$\lambda$3869/[OIII]$\lambda$5007=1.0$\pm$0.3 suggest ionization by shocks.}
\end{figure}

%blah

\begin{figure}
\includegraphics{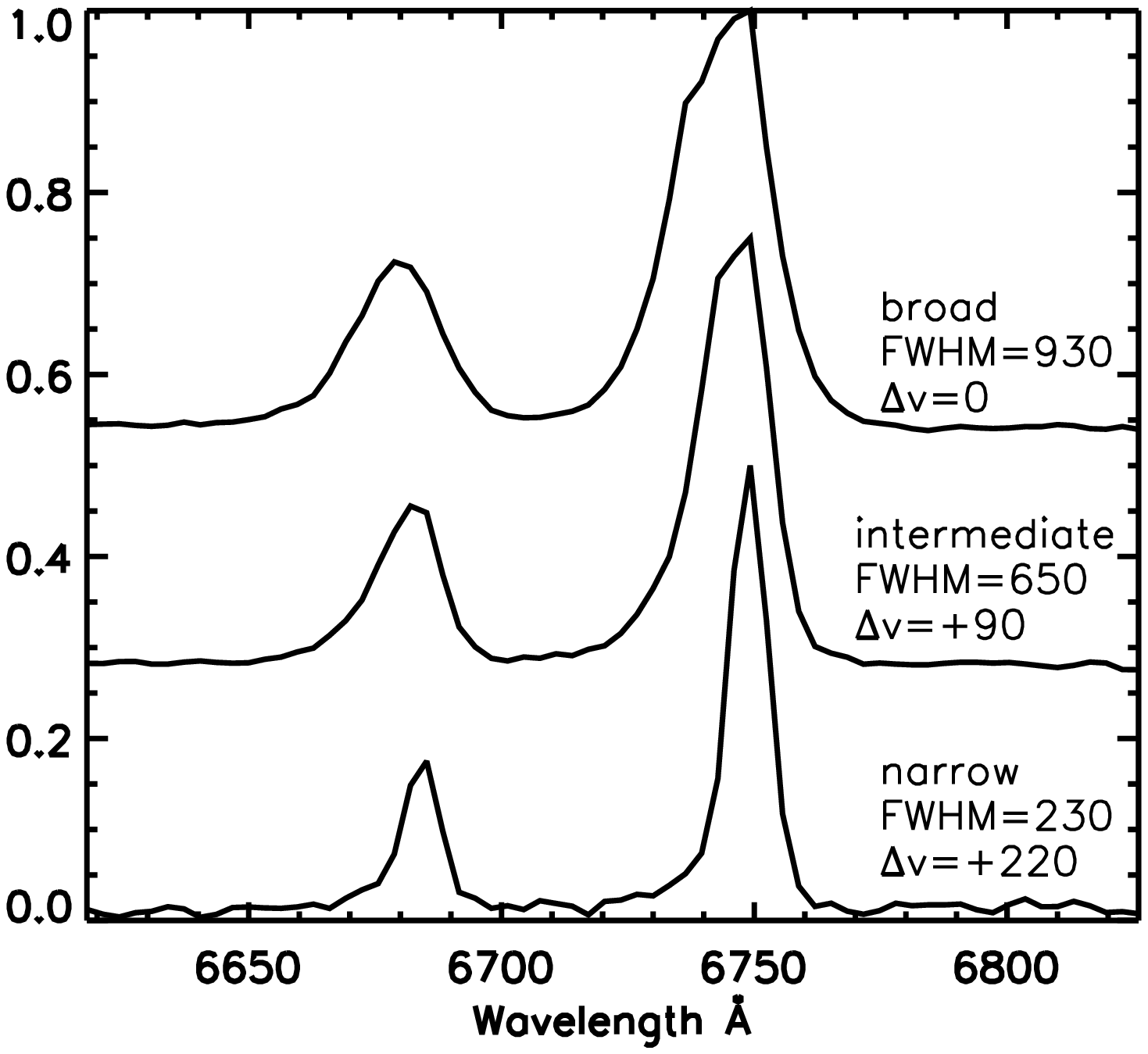}
\includegraphics{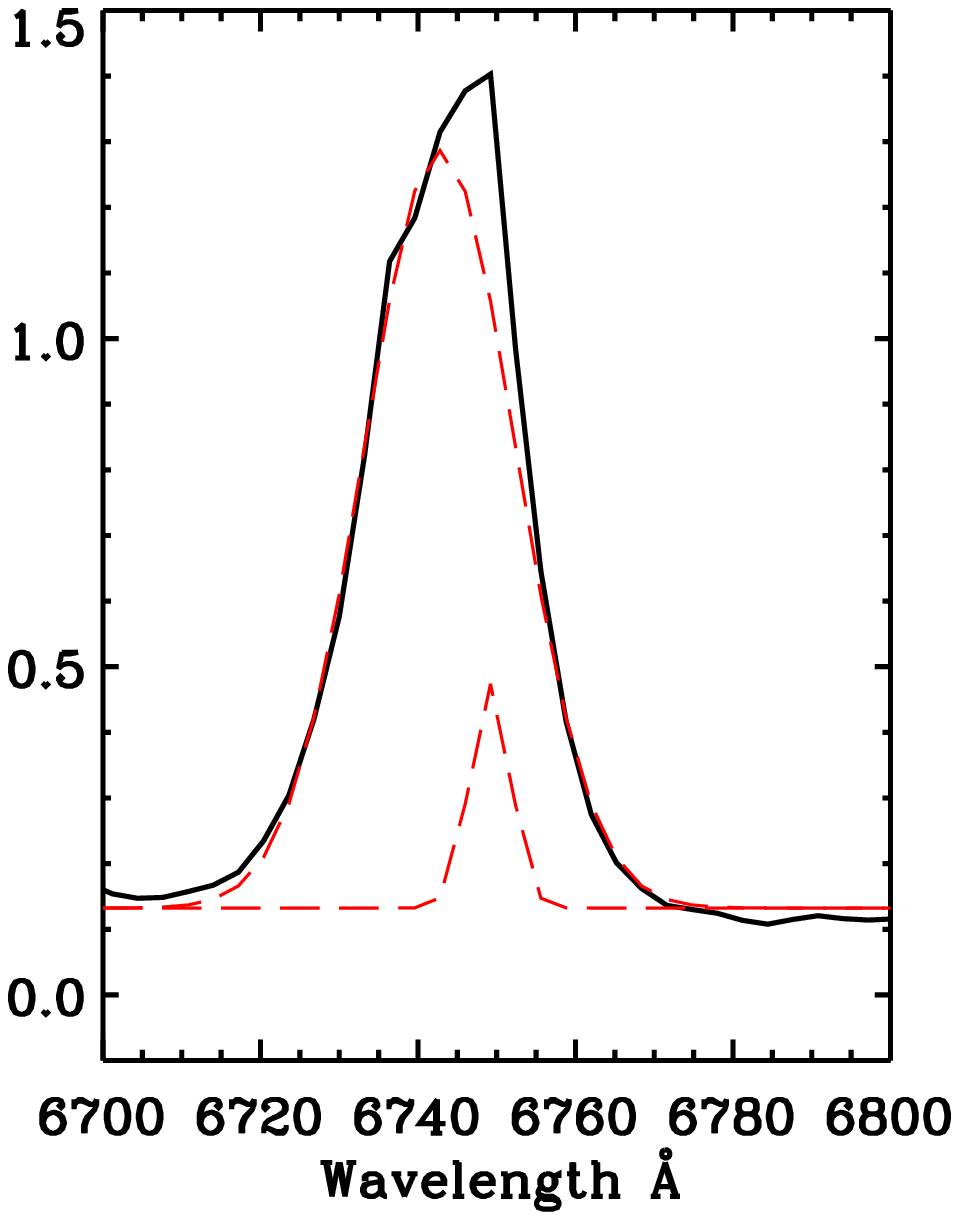}
\vspace{1.63in}\caption{{\it Left:} Velocity profiles of the [OIII] $\lambda\lambda$ 4959,5007 doublet in 1550+39, to illustrate the trend between FWHM and line of sight velocity $\Delta$v.  In this figure, an arbitrary flux scale was used.  The 'broad' aperture is the sum of the 6 fibres showing the highest FWHM, i.e., [-2:0,0:1].  The 'narrow' aperture corresponds to 6 fibres in the Northern region of the nebula, i.e., [-1:0,2:3], and includes the 3 fibres with the lowest FWHM in this object.  The 'intermediate' is the sum of fibres [-1:0,-1].  The values for $\Delta$v and FWHM are given in km s$^{-1}$.  {\it Right:} An illustration of how we 'detect' blue wings.  Here we show the observed [OIII] $\lambda$5007 line in the 'broad' aperture (solid smooth line) fitted using two Gaussians (dashed red lines).  The reduced $\chi^2$ for this fit is 18.7, which is substantially lower than the value 94.2 obtained from a single Gaussian fit (not shown).  In both panels, the flux density scale is in arbitrary units.}
\end{figure}

\begin{figure*}
\includegraphics{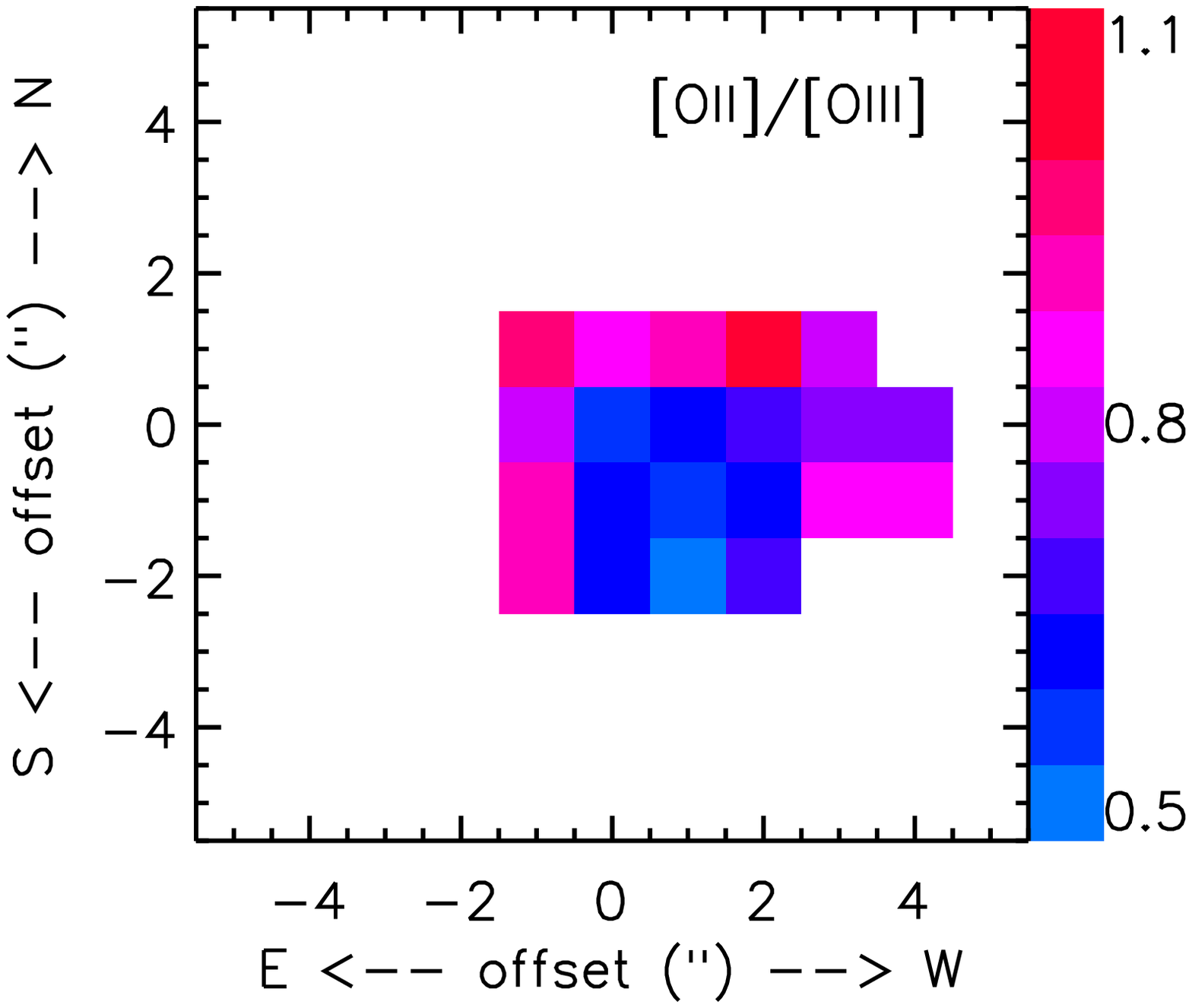}
\includegraphics{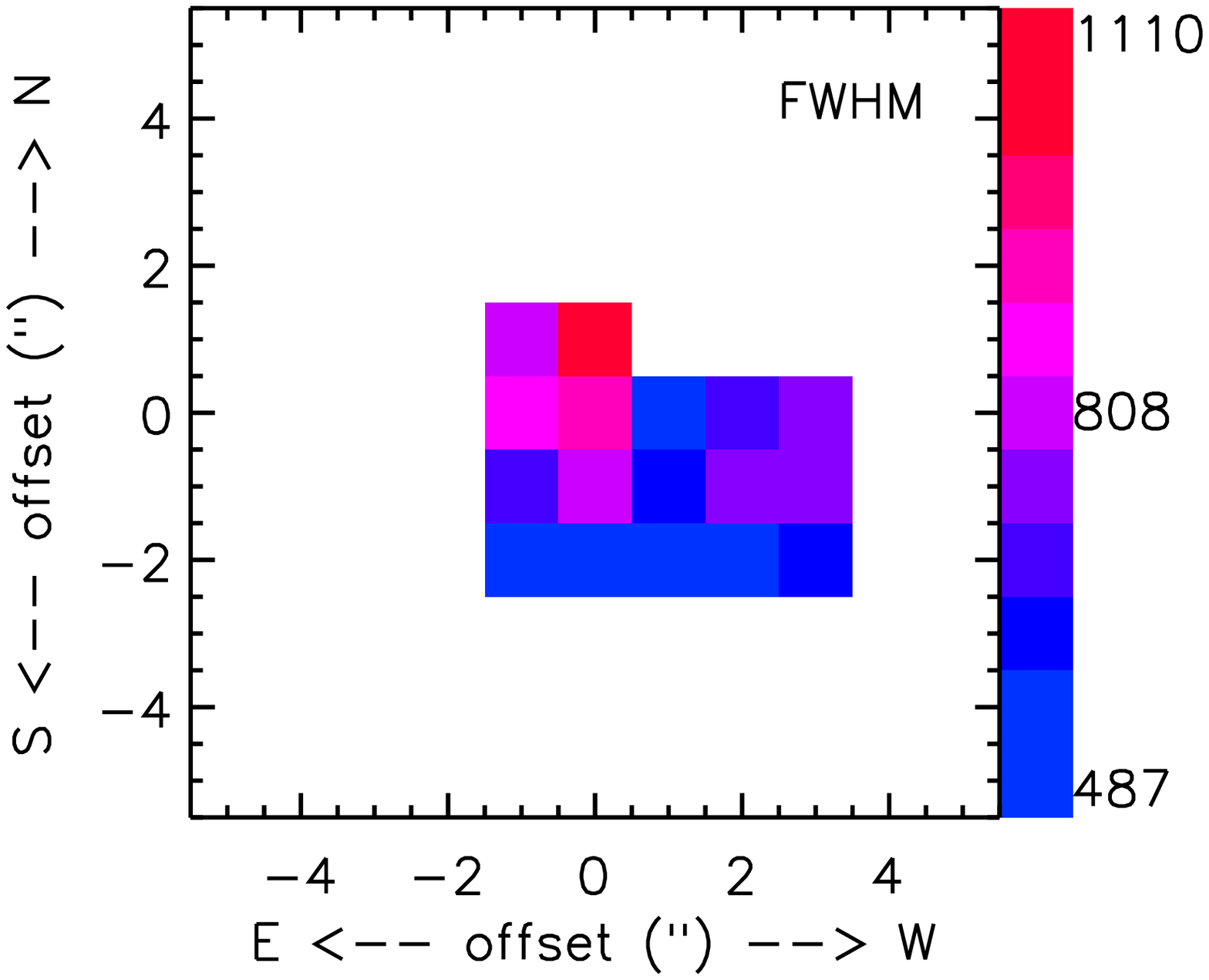}
\includegraphics{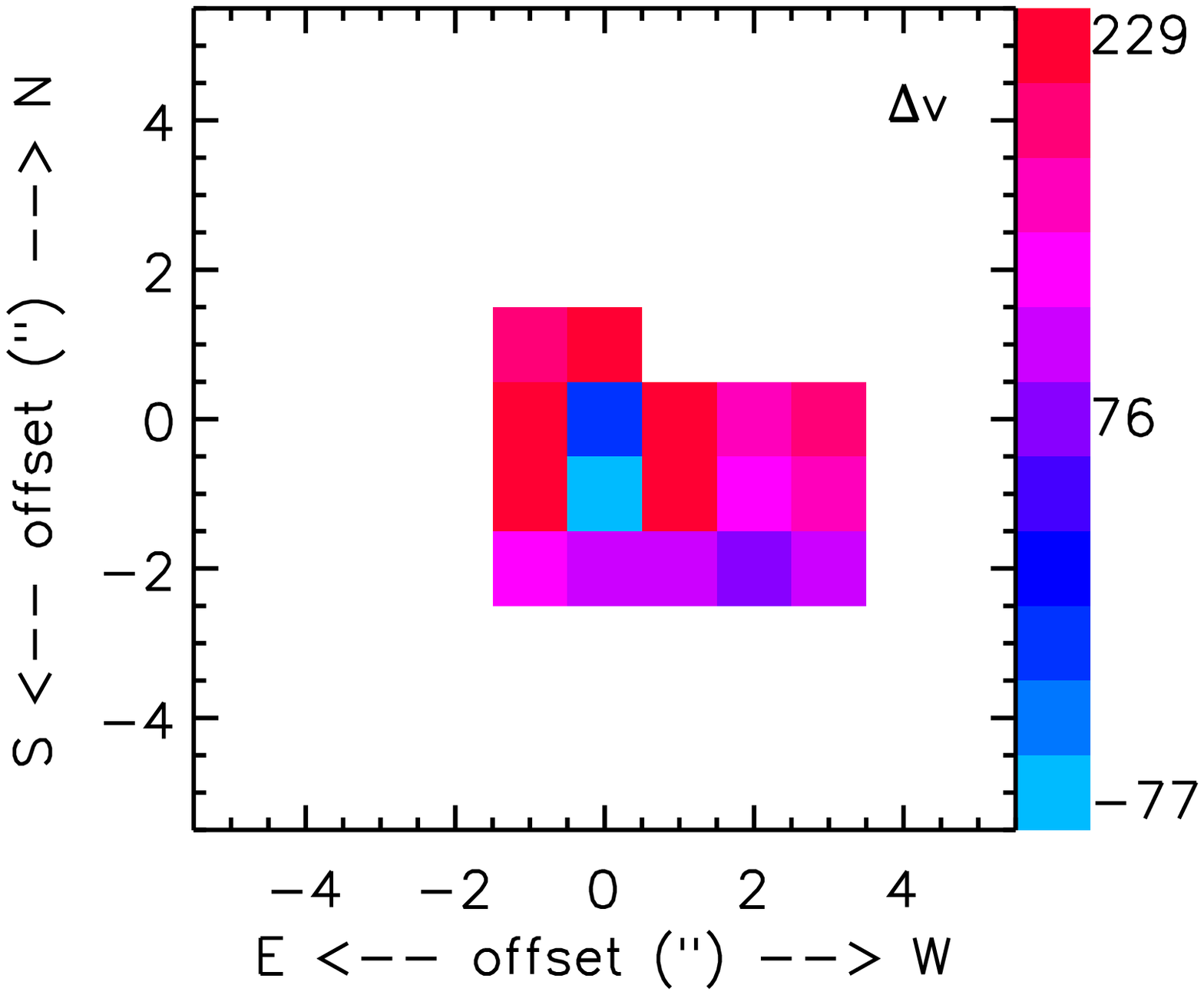}
\vspace{2.1in}\caption{Results from the datacube of SDSS J084041.08+383819.8.  From left to right: the [OII] $\lambda$3727 / [OIII] $\lambda\lambda$4959,5007 map; the [OIII] $\lambda$5007 FWHM map (km s$^{-1}$); map of the velocity shift of [OIII] $\lambda$5007 (km s$^{-1}$), adopting the velocity at the spatial zero as our fiducial velocity zero.  The range in [OII] $\lambda$3727 / [OIII] $\lambda\lambda$4959,5007,FWHM and velocity shift is typical of the spatially extended ionized gas associated with powerful active galaxies (see text). The total intensity map is similar to that shown in Fig. 1.}
\end{figure*}

\section{Summary and discussion}
In this letter we have presented new integral-field observations of 6 type-II QSOs at 0.3$\le$z$\le$0.4, obtained with PMAS on the 3.5m telescope at Calar Alto.  Two of these sources, SDSS J155059.30+395029.5 and SDSS J084041.08+383819.8, were found to be associated with extended ionized nebulae, which have minimum spatial extents of 27 kpc and 40 kpc, respectively.  Both nebulae exhibit varied kinematic properties: some regions show relatively quiescent kinematics, consistent with gravitational motion, while in other regions of the nebulae the kinematics are highly perturbed, with FWHM exceeding 1000 km s$^{-1}$.  We interpret the kinematic pattern in SDSS J155059.30+395029.5 as an outflow of $\sim$285 km s$^{-1}$ velocity which extends across an area of 20 kpc $\times$ 20 kpc, with a mass ejection rate of $\ga 2 M_{\odot}$yr$^{-1}$.  Plausible mechanisms to power this apparent outflow include a starburst, the radio jets, or the AGN radiation field.  

Furthermore, the nebula associated with SDSS J155059.30+395029.5 shows evidence for AGN photoionization, based on the detection of extended [NeV] emission, and evidence for ionization by shocks, based on a relatively high [NeIII]/[OIII] ratio (see e.g. Dopita \& Sutherland 1996).

Overall, the luminosities, sizes, excitation properties an kinematics of the two extended nebulae are consistent with those of radio galaxies and type I QSOs at similar redshifts (see e.g. Baum et al. 1988; Willott et al. 1999).  We suggest that AGN radiation powered outflows or star formation, either of which are able to explain the outflow from SDSS J155059.30+395029.5, might also be able to explain the similar kinematic properties shown by some nebulae associated with radio galaxies and quasars -- properties which are often assumed to be powered by radio jets.  

\section*{Acknowledgments}
We thank the anonymous referee for suggestions which helped improve this letter.  AH also thanks Brian McNamara for useful comments regarding the estimation of jet power from radio luminosity.  The work of MVM, RGD and EP has been funded with support from the Spanish MICINN grant AYA2007-64712.  AMS is supported by a UK STFC postdoctoral fellowship.

\bsp

\label{lastpage}

\end{document}